# IT Project Governance in Project Delivery: Key Processes, Activities, Roles and Responsibilities

Article by Godfred Yaw Koi-Akrofi
*Dean-School of Science and Technology, Dominion University College, Ghana*
*E-mail: godfred_akrofi@texilaconnect.com, g.koi-akrofi@duc.edu.gh, godfred_akrofi@yahoo.com*

### *Abstract*

*The general objective of this work was to contribute to the general body of knowledge and research work in the area of managing IT projects successfully. The office of Government commerce of the UK Government in conjunction with the National Audit came out with a guideline in 2007 which list out eight causes of project failures. Six out of the eight causes were attributed to governance issues (Aon, 2011). This shows clearly that governance activities in project management are really important, and failure to place premium on them can result in failure of projects. The research work was therefore intended to come out with IT governance frameworks that can help resolve the problem of failure of IT projects resulting from governance issues. Four IT program managers and eight IT project managers were talked to on a variety of IT project governance issues. The frameworks were developed based on a combination of literature, experts (IT project and program managers of the telecom industry in Ghana) input, observation, and so on. The governance frameworks depicts project management processes, project activities, project governance activities, roles and responsibilities of key stakeholders, key milestones, approval bodies, signatures, and so on, to ensure successful delivery of IT projects.*

## Introduction

IT projects delivery entails a lot of activities from the beginning to the end; the project manager, his/her team, and all those in the project environment (stakeholders) must engage themselves in a lot of activities to ensure the successful delivery of the project. These activities must be well followed and coordinated to ensure a good flow for the project to be delivered successfully. IT projects unlike other projects are unique because in most cases, it is a combination of systems development processes and traditional project management processes. In some other projects, you may not need to do requirements discovery, gathering, and analysis, but these are strongly required in IT projects.

Project governance is key to the successful delivery of projects. Project governance deals with the facets or parts of governance that has a relation with ensuring the efficacy of projects. Project governance has elements of both corporate (or organisational) governance and specific or exact project management systems (HM Treasury, 2007). Corporate governance has a relation with the answerabilities and duties for how performance is managed in an enterprise (Aon, 2011). Governance originates from the Greek, to steer, or to guide. The word governance is linked with words like government, governing and control (Klakegg et al., 2008). In the organizational setting, governance provides a context or background for ethical decision making and managerial action in an organization that is depended on pellucidity, accountability, and defined roles (Muller, 2009).

There are two different views about governance. The first class of people suggest that each unit within an organization needs its own governance. Some of these different types of governance include papers "on IT governance": (Marnewick and Labuschagne, 2011; Martin and Gregor, 2006; Sharma et al., 2009; Willson and Pollard, 2012); "on knowledge governance": (Ghosh et al., 2012; Pemsel and Müller, 2012); "on network governance": (Klijn, 2008; Sørensen, 2002); "on public governance": (Du and Yin, 2010;

Klakegg et al., 2008; Williams et al., 2010); and "on project governance": (Abednego and Ogunlana, 2006; Miller and Hobbs, 2005; Winch, 2001). This view of governance seems to be very popular among IT managers, project managers, officials within government departments, and academics who work solely within these disciplines. Their take is that governance is a function of management, and that each governance practice operates independently from the other and in practice, they do not mix or there is no integration. This is the view of the author, and the research is based on this view. The author believes that each unit/discipline within an organization should have its own governance for effective tracking of performance. The second view is championed by organizations such as the OECD (OECD, 2004), various Institutes of Directors (e.g. Australian Institute of Company Directors, 2010; Institute of Directors Southern Africa, 2009) and the agencies responsible for governing stock exchanges. In this view, governance is a single process with different facets or sides. An example, they argue, is a governing organization under various themes or subjects: governing relationships, governing change, governing the organization's people, financial governance, viability and sustainability.

IT governance refers to the management framework within which IT project, program and/or portfolio decisions are made (https://uwaterloo.ca/it-portfolio-management/methodologies/roles-and-responsibilities/it-governance). According to Wilson and Connel, n.d, project governance is a framework that sets out the structure, resources, communication, reporting and monitoring systems to manage a project in consonance with an organization's corporate or strategic vision. They continued to say that project governance is the framework which certifies that the project has been rightly thought of and is being implemented or done in accordance with best project management practice and within the broader framework of the organization, and that effective project governance is about making sure that projects deliver the value expected of them. An appropriate governance framework helps save money by making sure that all spending is fitting for the risks being tackled. Project governance is far from micro-management, rather it is about setting the terms of reference and operating framework, defining the boundaries and making sure that planning and execution are carried out in a way which ensures that the project delivers benefits (https://www.brookes.ac.uk/services/hr/project/survival/governance.html). Project governance is about guiding and monitoring the process of translating investment decisions into value for the organisation, delivering the estimated and predicted benefits – the business outcomes and benefits to envisioned beneficiaries or recipients (Department of Treasury and Finance, 2012). According to the department of treasury, 2012, the four key principles for effective project governance are:

1. Establish a single point of overall accountability or responsibility.
2. Ownership of service delivery determines project ownership.
3. Stakeholder management must be separated from project decision making.
4. Distinguish or differentiate between project governance and organisational structures.

The following are a list of the role of governance in project management:

1. Developing and managing processes
2. Ensuring guidelines and procedures
3. Managing risk
4. Managing change proposals
5. Monitoring and reporting on governance processes to ensure adherence to standards
6. Ensuring projects are on track to meet expected outcomes
7. Ensuring appropriate levels of resources are assigned
8. Reporting on or based on definite levels
9. Providing an overview or map of agreed process for completing a project
10. Observing or following strictly guidelines and procedures
11. Full clarity or transparency
12. Project structure
13. Objective oversight (Wilson & Connel, n.d),

According to HM Treasury, the main activities of project governance relate to:

- How the programme is directed;
- How projects are owned and the sponsorship;
- How to ensure the efficiency or success of project management roles; and
- How to report and disclose (which includes liaising with stakeholders).

(HM Treasury, 2007).

The four areas mentioned above cut across all the phases of the project life cycle. For the purposes of this research work, and combining the project management and systems development activities, we conclude on four phases of the project life cycle namely, Project Initiation, Project Planning, Project Delivery/Implementation, monitoring and controlling, and Project Closure. At every stage of life cycle, there are a number of project activities, as well as governance activities.

Apart from contributing to the general body of knowledge and research work in the area of IT projects, this research work is aimed at coming out with IT governance frameworks that will depict project management processes, project activities, project governance activities, roles and responsibilities of key stakeholders, key milestones, and so on, to ensure successful delivery of IT projects.

The office of Government commerce of the UK Government in conjunction with the National Audit came out with a guideline in 2007 which list out eight causes of project failures. Six out of the eight causes were attributed to governance issues (Aon, 2011). This shows clearly that governance activities in project management are really important, and failure to place premium on them can result in failure of projects. The research work intends to come out with IT governance frameworks that can help resolve the problem of failure of IT projects resulting from governance issues.

## Materials and methods

The frameworks were developed based on a combination of literature, experts (IT project and program managers of the telecom industry in Ghana) input through rigorous interviews and discussions, project documentation of selected firms/companies in the telecom industry in Ghana (firms who have a structured system for project management practice), observation, and so on. The telecom industry in Ghana was used as a case study. Four IT program managers and eight IT project managers were talked to on a variety of IT project governance issues.

## Results and discussions

We start the discussion with some few governance frameworks in literature. Aon (2011) presents a project delivery framework showing all the phases of the project life cycle, as well as the deliverables of each phase. Table 1 below gives a summary of it.

**Table 1.** Summary of project delivery framework

| **Project Lifecycle** | **Strategic Assessment** | **Project Initiation** | **Project Planning** | **Project Execution** | **Project Closure: Handover** | **Project Closure: Operations** |
|---|---|---|---|---|---|---|
| **Key Governance Documents** | *Strategic plan at programme or organization level* | | | | *User acceptance*<br>*Project handover plan*<br>*Project final report with KPI* | |
| **Key Governance activities** | *Multi-year strategic planning of programmes* | | | | | |
| | *and projects* | | | | | |
| **Major Milestones** | *Plan Approved* | *Business Justification Project feasibility* | *Project Governance Committee formed* | *Project commenced* | *Project handover to operations Project management team dissolved* | *Operational review completed Project governance committee dissolved Post project review* |
| **Key Approvals** | | *Strategic decision to proceed* | | *Resources committed* | *Approve project completion* | |
| **Key Checkpoints** | *Review of strategic fits of projects at programme level* | *Business case of the project* | *Delivery Strategy* | *Investment Decision* | *Readiness for service* | *Operational review and benefits realisation* |

Source: Aon (2011).

From Table 1 we see five stages of the project life cycle, and for each stage, what is required in terms of key governance documents, key governance activities, major milestones, key approvals and key checkpoints. The governance activities are spelt out in the governance documents for each phase. We also have stages where there are key approvals, and for that matter things that must be checked to ensure satisfaction of governance activities before the next stage is allowed. This rigorous framework is necessary to ensure nothing is left out in terms of governance activities to ensure successful delivery of projects.

In Ralf Muller's book PM Concepts, he presents a framework for project governance that is worth discussing. In his book, he talks about three incremental project management governance steps. The first step is made up of the methodology, Steering Committee, and Reviews and Audit. At this step, the constraints are "very strong technology focus", and the enablers are "Better Project results needed". The second step is made up of Certification, Project Management Office (PMO) and Project Support Office (PSO), and Mentoring. The constraints here is "Strong project control focus", and the enablers, "Role and career for project managers". The last and third step is made up of advanced training, Bench Marking, and Maturity models. The constraints are "Resistance to change existing practices", and the enablers, "project management as strategic capability".

Too and Weaver (2014) also came out with a model for project governance. Theirs is a little bit different from Ralf Muller, in that, they looked at four levels of management namely: project/program management, senior management, executive management, and Governing board. The project/program management and the senior management oversee the project delivery system and the management system; and the executive management and the governing board oversee the governance system and the management system. The four key interrelated elements to support effective governance of projects and programs according to this framework are Portfolio Management; Project Sponsors; Strategic PMO and Effective Projects and Programs Management. According to Too and Weaver (2014), project governance in a multi-project environment has two main functions. The first function is a decision about which projects the organization should approve, fund and support. Management is then made aware of these decisions for implementation. The essential yields from this piece of the project governance framework are: particulars or specifications about the rights and duties of key members or players in the projects (partners or stakeholders); meanings of (and agreement for) standards and strategies for deciding; advancement of the vital framework/model for choice of the "right" projects and projects to embrace – including a reasonable comprehension of what "right" means for every organization; lastly components for the productive and compelling utilization of assets or resources. The second function of the project governance system is the oversight and assurance. These capacities or functions include: concurring the current strategic plan (in conjunction with top or executive management) and how the projects affirmed within that strategy add to the organization strategic targets or objectives; correction of the strategic plan because of evolving conditions; monitoring performance of the projects within the strategic plan and the stewardship (effective management) of assets/resources connected to these projects; correspondence of these affirmations suitable outside partners or stakeholders, the organization's owners, and the more extensive stakeholder group (including regulatory authorities).

The main focus of the interviews of the IT project and program managers was to determine the project activities and governance activities for each process group or phase of the project management life cycle (PMLC), the roles and responsibilities of stakeholders, the major milestones and deliverables within the life cycle, the major issues of governance, and so on. A typical systems development will employ the systems development life cycle or the waterfall model. The waterfall model has the processes as analysis, design, coding/construction/implementation, testing, deployment, and maintenance. Matching this with the Project management Institute's (PMI's) PMLC, we have Initiating to Analysis; Planning to Design; Executing, monitoring and controlling to coding/construction/implementation; Closing to Testing and Deployment, and then the last stage of waterfall which is maintenance. Basically, from the interviews, there are four main categories which need to be dealt with in terms of frameworks for project governance, and these are Roles and Responsibilities: Steering committee, Project Team, Stakeholders; Tools and methodologies: Project management, Quality management, Financial management, Change management, and so on; Control mechanisms: Financial, Deliverable management, Communication, Human Resources, Confidentiality; Risk management: Financial, Change Management, Delivery, Other (Legislative; Technology; Political; Organizational culture; Human; and so on.).

A project is first conceived as an idea, either from Strategy, Competition or Operations. The idea goes through a thought process and is then aligned with Company Vision, Objectives and Values. If idea passes initial Business Case assessment, approval is given by Business to embark on detailed feasibility of project to ascertain what it takes to deliver project outcome, including related cost, associated risk, people required, etc. A Feasibility report will summarise all findings and will be issued for review and approval by Decision maker. Business Case and Feasibility Report need to be approved together before Implementation can start.

For most start of Implementation, Kick-off meeting is needed to bring stakeholders together to align on the Objectives of the project plus setting up of ground rules/protocols including reporting, communication plan, escalation path, resource matrix and any other team responsibility. During implementation, communication plan, Reporting and Escalation path tools are used as guidelines to manage how information on project is shared. After deployment, project enters Testing phase where acceptance criteria is matched to product delivered. There are varying types of test required in projects. The critical must-do test cases include system test, User acceptance test, regression test and for some cases dry-runs or simulation test. Each type of test is expected to pass pre-defined acceptance criteria and have scores within the customer preferred thresholds. When all tests required are completed and results are accepted by Senior Users, Customers and Operations, the project is then declared as READY for Service (RFS). Once RFS is achieved, project would then be prepared for live environment and Go-Live. New systems introduced in Business and are being used for first time will normally go through a simple switch-on then live environment would be activated for use. On the other hand, for a system swap, which is new system replacing existing, a carefully planned migration strategy would be required. This would be necessary to ensure minimal database errors, missing data, and duplicated data, among others.

Some category of projects require no change in business processes and customer engagements, it's just a seamless transition from project to live environment, and for such, project closure can commence immediately after go-live. On the Other hand if a Project delivers a new product to Market, or delivers a major change in Business operations, then a Product Formal Launch becomes necessary. With product launch, Product is out-doored, customers are educated through ads and various communication channels. Product launch gives publicity and awareness of product to both internal and external users to help drive early benefit realisation referred to in the business case.

Project hand-over to Operations' arm of a Business happens after go-live and launch. Operations are responsible for day-to-day management of product until end of product life. PM will have to deliver to Operations per agreed hand-over check list. Project Closure is seen as the last milestone of a project, after which the project is deemed as completed. This includes submission of hand-over documents, disbanding of team and resources, capturing of lessons learnt for project library, and organising of final project closure review. It is recommended as best practice that the business case of a project be reviewed after a minimum period of six months of post implementation, to assess benefit realisation. It is also a common practice to include a period of stabilization for turn-key projects, which allow room for suppliers to monitor deployed system over a time before final hand-over to customer and this can also serve as an opportunity for hand-holding and up skill of customer.

Table 2 below shows a framework on project governance based on the perspectives of the interviewers, who are expects in the IT projects field.

**Table 2.** Project governance framework

| **Traditional Process Group** | **Waterfall (for systems development)** | **Governance life cycle process** | **Key project activities** | **Key Governance activities** | **Key deliverables** | **Approval body/signatures** |
|---|---|---|---|---|---|---|
| Initiating | Analysis | Idea/thought/ notion and Concept Stages | •Develop business case<br>•Undertake a feasibility study<br>•Establish the project charter<br>•Assess risks<br>•Determine scope<br>•Draft plan (budget, schedule, team)<br>•Set up the project office<br>•Appoint the project team<br>•Perform a phase review | •Assess strategic fit<br>•Assess risks and return<br>•Approve project<br>•Appoint project governance team<br>•Determine expected outcome/benefits | *Project Request Form /Idea Summary<br>*Proposal<br>-Business case, including cost/benefit analysis<br>-Initial budget requirements & funding source identification<br>-Preliminary Security review<br>*Contract Terms & Conditions (if required)<br>*Business Requirements<br>*Project Charter | *Project Sponsor (Business Unit (BU) Head or Tech Functional Heads)<br>*Technical Lead<br>*Project Manager<br>*Programme Manager |
| **Traditional Process Group** | **Waterfall (for systems development)** | **Governance life cycle process** | **Key project activities** | **Key Governance activities** | **Key deliverables** | **Approval body/signatures** |
| Planning | Design | Feasibility Solution Design | •Create a Project Plan<br>•Create a Resource Plan<br>•Create a Financial Plan<br>•Create a Quality | *Approve terms of reference of project governance team<br>*Approve project objectives, plans, milestones, and Key Performance Indicators (KPIs)<br>*Approve project | *Detailed project plans<br>*Assembled project team | Project Control Board (PCB) for approval of mandate and start of implementation, and signatures as:<br>*BU Head / Sponsor<br>*Project Manager |
| | | | Plan<br>•Create a Risk Plan<br>•Create an Acceptance Plan<br>•Create a Communications Plan<br>•Create a Procurement Plan<br>•Contract the Suppliers<br>•Form team<br>•Plan deliverables<br>•Procure goods and services | management team | | *Project Management Office (PMO) Head<br>*Solutions Dev. Head For PFA (Project Funding Agreement) we have approval body as senior management team (SMT) and signatures as:<br>*Project Manager<br>*BU Head / Sponsor<br>*Heads PMO, Finance<br>*Chief Technical Officer (CTO),<br>*Chief Financial Officer (CFO), Chief Executive Officer (CEO). |
| **Traditional Process Group** | **Waterfall (for systems development)** | **Governance life cycle process** | **Key project activities** | **Key Governance activities** | **Key deliverables** | **Approval body/signatures** |
| Executing/ Monitoring and Controlling | Coding/ Construction/ Implementation | Implementation | *Manage project<br>*Manage stakeholders<br>*Manage risks and changes<br>*Monitor and control<br>*Report | *Monitor project progress against KPIs<br>*Make high level project decisions<br>*Determine strategic change to the projects<br>*Resolve key issues<br>*Monitor risks | Product, service or system | Approval body: None<br>Signatures:<br>*Project Manager<br>*Head of Service Assurance<br>*BU Head |
| **Traditional** | **Waterfall (for systems** | **Governance life cycle** | **Key project activities** | **Key Governance activities** | **Key deliverables** | **Approval body/signatures** |

| **Process Group** | **development )** | **process** | | | | |
|---|---|---|---|---|---|---|
| Closure | Testing and Deployment | *Test<br>*Launch<br>*Observation period<br>*Business as usual (BAU) | *Close project<br>*Review project performance | *Assess and approve transition from project team to operational team<br>*Assess realization of expected benefits | Product/<br>Service/<br>System ready for use | **Testing**<br>Approval body: Offline PCB<br>Signatures:<br>*BU Head / Sponsor<br>*Head Com. Finance<br>*CTO<br>*PMO ( Project Manager)<br>**Closure**<br>Approval body: PCB/Offline<br>Signatures:<br>*Project Sponsor<br>*Vendor<br>*Project Manager<br>*Programme Manager |

**Source:** Author (Godfred Yaw Koi-Akrofi)

Figure 1 below shows the basic process of delivering IT projects based on focussed interviews with project and program managers in the Telecommunications industry in Ghana

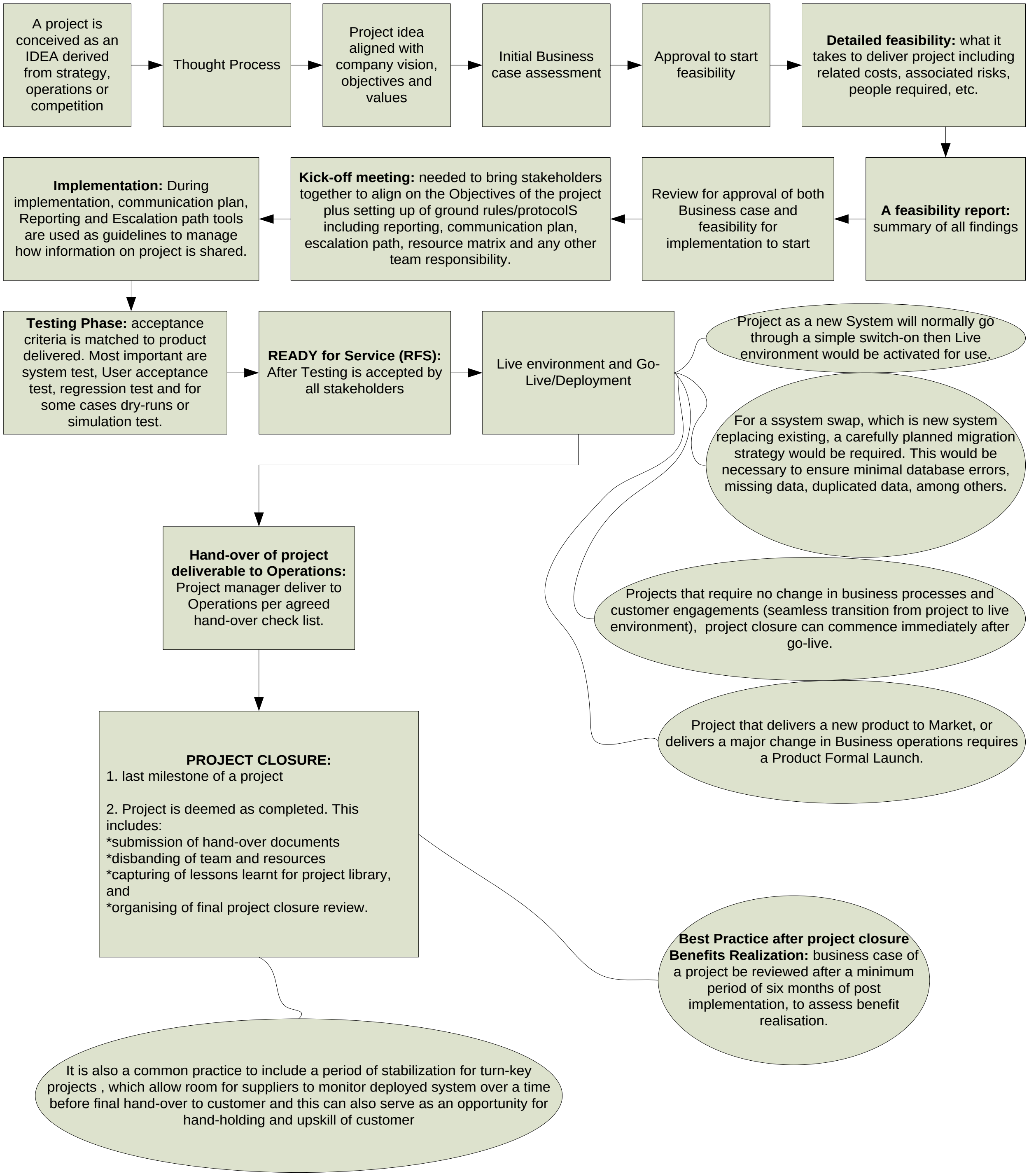

**Figure 1.** Project delivery and governance framework

## Conclusion

From table 2 and figure 1 and the discussions with the IT project and program managers, the following conclusions are made:

1. The initial stages of IT projects are very crucial and must be painstakingly executed to ensure success at the end of the day. This is achieved by following the right processes and procedures interspersed with approvals before the next stage is take; these are all governance activities.
2. Project governance is crucial throughout the entire process to ensure a single point of overall accountability or responsibility, ownership of service/product delivery, direction of the project in general, total sponsorship of project, efficiency or success of project management roles; and to report and disclose (which includes liaising with stakeholders).
3. Project governance activities are totally different from project activities and must not be confused, but are integral to the success of a project
4. Project delivery without project governance breeds corruption, compromise on standards and quality, ineffectiveness, delayed projects, lack of proper communication and reporting, project funding problems, ownership and direction issues, and so on.

## Acknowledgement

I would like to express my sincere thanks to my wife Joyce Koi-Akrofi who served as a main resource right from the beginning to the end. My thanks also go to all the program and project managers who out of their tight schedule could still manage to attend to my numerous calls for meetings, and so on. I appreciate all who contributed in one way or the other for the success of this paper.